\documentclass[aps,prb,twocolumn]{revtex4-1} 
\pdfoutput=1
\usepackage{amsmath,amssymb,amsfonts,upgreek}
\usepackage{color}
\usepackage{graphicx}
\usepackage{setspace}
\usepackage{bm}
\usepackage{comment}
\usepackage{hyperref}

\begin{document}

\title{Interplay between breathing mode distortion and magnetic order
  in rare-earth nickelates \textit{R}NiO$_3$ within DFT+$U$}

\author{Alexander Hampel}
\affiliation{Materials Theory, ETH Z\"u{}rich, Wolfgang-Pauli-Strasse 27, 8093 Z\"u{}rich, Switzerland}
\author{Claude Ederer}
\affiliation{Materials Theory, ETH Z\"u{}rich, Wolfgang-Pauli-Strasse 27, 8093 Z\"u{}rich, Switzerland}

\date{\today}

\begin{abstract}
We present a systematic density functional theory (DFT) plus Hubbard
$U$ study of structural trends and the stability of different
magnetically ordered states across the rare-earth nickelate series,
\textit{R}NiO$_3$, with $R$ from Lu to La. In particular, we
investigate how the magnetic order, the change of the rare-earth ion,
and the Hubbard interaction $U$ are affecting the bond-length
disproportionation between the nickel sites. Our results show that
structural parameters can be obtained that are in very good agreement
with present experimental data and that DFT+$U$ is in principle able
to capture the most important structural trends across the nickelate
series. However, the amplitude of the bond-length disproportionation
depends very strongly on the specific value used for the Hubbard $U$
parameter and also on the type of magnetic order imposed in the
calculation. Regarding the relative stability of different magnetic
orderings, a realistic antiferromagnetic order, consistent with the
experimental observations, is favored for small $U$ values and
becomes more and more favorable compared to the ferromagnetic state
towards the end of the series (i.e., towards $R=$Pr). Nevertheless, it
seems that the stability of the ferromagnetic state is generally
overestimated within the DFT+$U$ calculations. Our work provides a
profound starting point for more detailed experimental investigations,
and also for future studies using more advanced computational
techniques such as, e.g., DFT combined with dynamical mean-field
theory.
\end{abstract}

\maketitle

\section{Introduction}

Materials that are located at the crossover between itinerant and
localized electronic behavior often exhibit rich phase diagrams,
including different forms of electronic order (charge, orbital,
magnetic) and metal-insulator
transitions.~\cite{RevModPhys.70.1039,Dagotto/Tokura:2008} Moreover,
exotic properties such as non-Fermi liquid
behavior,~\cite{Stewart:2001ks} high-temperature
superconductivity,~\cite{Dagotto:1994} or colossal
magnetoresistance~\cite{Dagotto/Hotta/Moreo:2001} can typically be
found in this regime, and in many cases a strong coupling between
electronic and lattice degrees of freedom, such as, e.g., the
Jahn-Teller effect,~\cite{Kanamori:1960,Dagotto/Hotta/Moreo:2001} can
be observed.

\begin{figure}[t]
\centering \includegraphics[width=1.0\columnwidth]{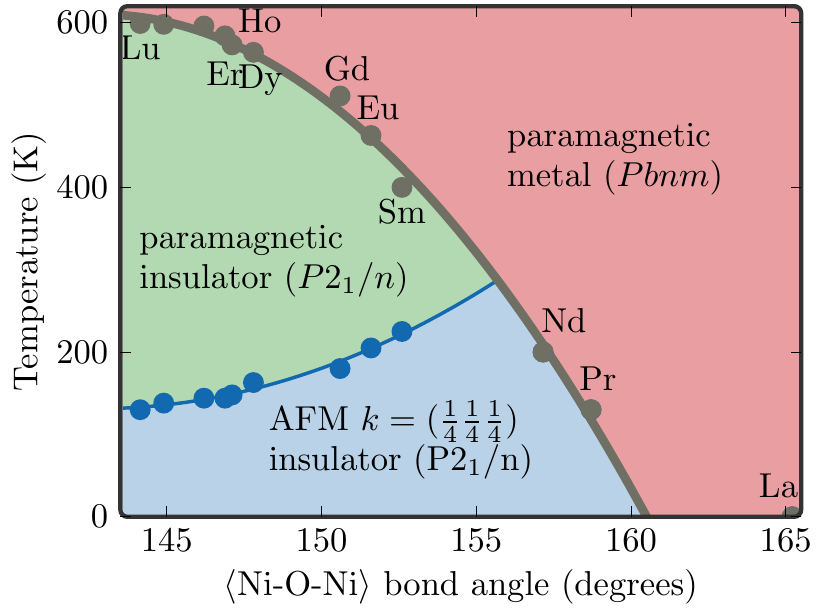}
\caption{Phase diagram of the rare-earth nickelates. Transition
  temperatures are shown as a function of the average
  $\langle$Ni-O-Ni$\rangle$ bond angle. Three different phases can be
  distinguished: i) paramagnetic metal with $Pbnm$ symmetry (red), ii)
  paramagnetic insulator with $P2_1/n$ symmetry (green), and iii)
  antiferromagnetic (AFM) insulator with $P2_1/n$ symmetry
  (blue).~\cite{footnote1}}
\label{fig:phase-diagram}
\end{figure}

An interesting example to study the crossover between localized and
itinerant electronic behavior is found in the series of
perovskite-structure rare-earth nickelates, \textit{R}NiO$_3$, where
\textit{R} can be any rare-earth ion ranging from Lu to
La.~\cite{Medarde:1997vt,Catalan:2008ew} All members of this series
(except LaNiO$_3$) exhibit a metal-insulator transition (MIT) as a
function of temperature, which is accompanied by a structural
distortion that lowers the space group symmetry of the crystal
structure from orthorhombic $Pbnm$ in the high temperature metallic
phase to monoclinic $P2_1/n$ in the low temperature insulating
phase.~\cite{Alonso_et_al:1999,Alonso:1999gk,Alonso:2000fz,Alonso:2001bs}
In addition, all systems (except LaNiO$_3$) order
antiferromagnetically at low
temperatures.~\cite{Garcia-Munoz_Rodriguez-Carvajal_Lacorre:1992,Medarde:1997vt}
The corresponding phase diagram (based on experimental data taken from
Refs.~\onlinecite{Medarde:1997vt,Alonso:1995cl,Alonso:2000fz,Alonso:2001bs,Alonso:1999gk,Munoz:2009go,Alonso:2013gt})
is depicted in Fig.~\ref{fig:phase-diagram}, where the temperature
dependence of the phase boundaries is shown as a function of the average
$\langle$Ni-O-Ni$\rangle$ bond angle. It can be seen that the
transition temperature for the MIT, $T_\text{MIT}$, decreases
monotonously with increasing $\langle$Ni-O-Ni$\rangle$ bond angle,
whereas the antiferromagnetic (AFM) transition temperature,
$T_\text{N}$, increases up to $R$=Sm but then becomes identical to
$T_\text{MIT}$. Thus, for $R$ from Lu to Sm, the AFM transition occurs
at lower temperatures than the MIT, whereas for $R$=Nd and Pr,
$T_\text{N}$ coincides with $T_\text{MIT}$. In contrast, LaNiO$_3$ is
a paramagnetic metal at all temperatures, and exhibits a slightly
different, rhombohedrally-distorted perovskite structure with
$R\bar{3}c$ symmetry.~\cite{GarciaMunoz:1992if}

The rare-earth nickelates allow us to study the transition from itinerant
paramagnetic behavior to a localized AFM state in a quasicontinuous
fashion using simple stoichiometric bulk systems, i.e., without the
need to introduce dopants or substitutional atoms. Moreover, the
nickelates are also highly tunable by pressure, strain,
electromagnetic fields, or doping, and are potentially multiferroic
(see, e.g.,
Refs.~\onlinecite{Catalan:2008ew,He:2015ey,ADMA:ADMA201003241}). Consequently,
the perovskite rare-earth nickelates have received considerable
attention during recent
decades.~\cite{Medarde:1997vt,Catalan:2008ew,Freeland:2015iw,Middey:2016jc}

\begin{figure}[t]
\begin{center}
  \includegraphics[width=0.75\columnwidth]{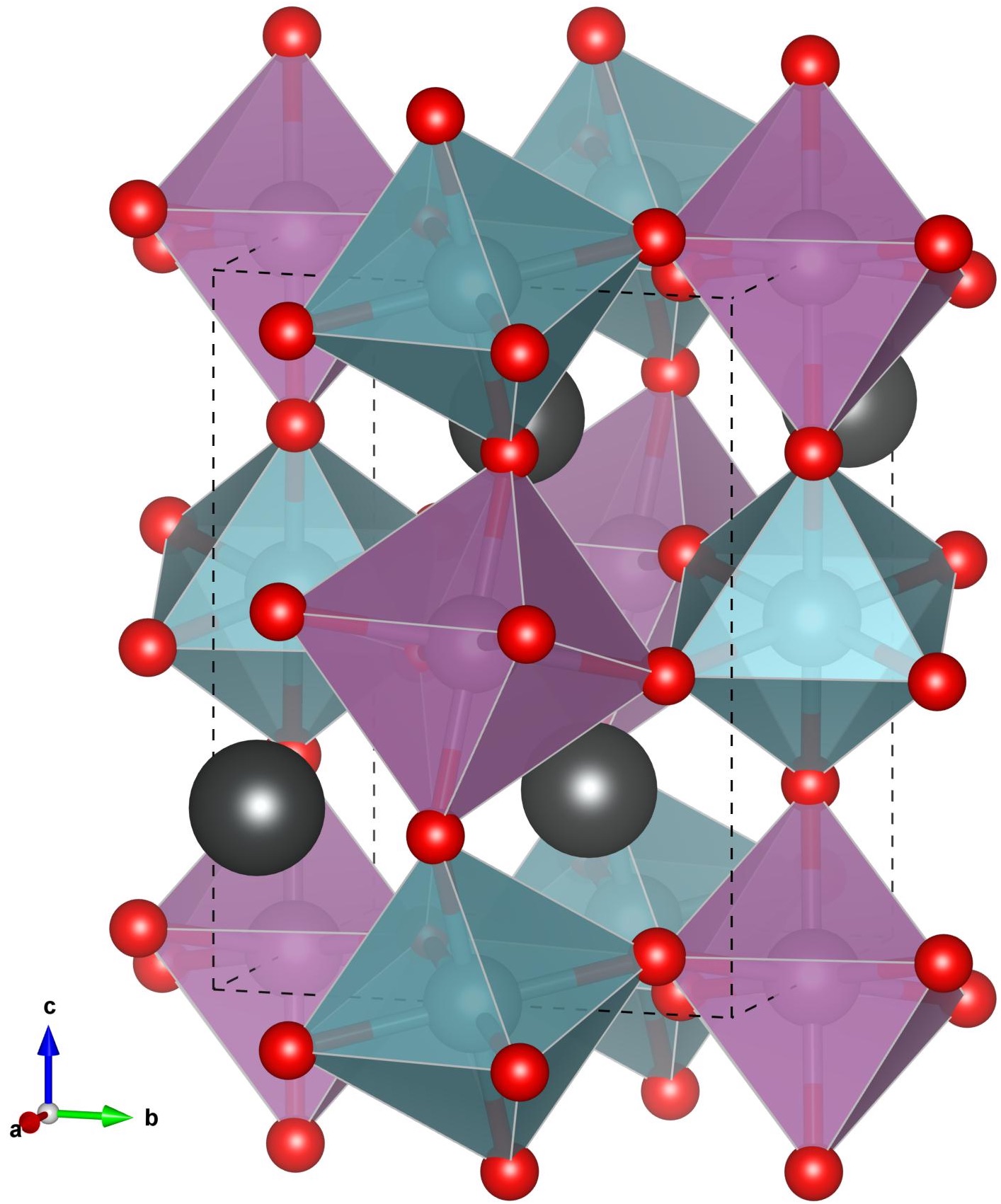}
\end{center}
\caption{Experimentally observed $P2_1/n$ crystal structure of
  LuNiO$_3$ at $60$~K below the MIT.~\cite{Alonso:2001bs} LB octahedra
  around the Ni sites are shaded in purple, SB octahedra are shaded in
  green, black spheres represent Lu, and red spheres represent O. The
  volumes of the bond-disproportionated NiO$_6$ octahedra differ by
  $\sim$$12$\,\%. The crystal structure is visualized using VESTA
  \cite{Momma:2008eu}.
\label{fig:exp-struc}}
\end{figure}

\begin{figure}[t]
\begin{center}
\includegraphics[width=0.75\columnwidth]{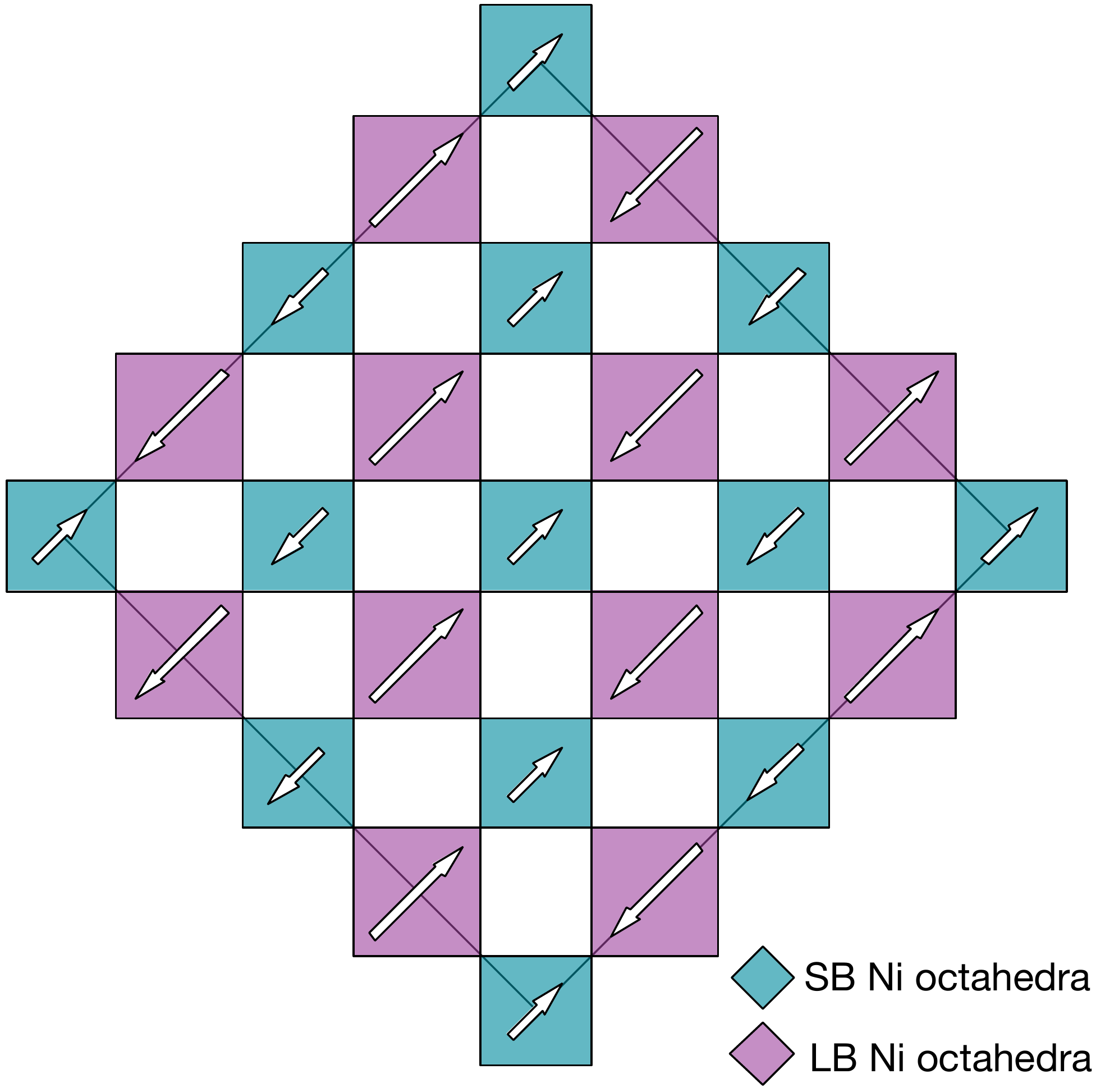}
\end{center}
\caption{Schematic depiction of a $[001]$-type layer of
  bond-disproportionated NiO$_6$ octahedra. Additionally, one of the
  experimentally suggested magnetic orderings
  ($T$-AFM)~\cite{Garcia-Munoz_Rodriguez-Carvajal_Lacorre:1992,Medarde:1997vt}
  is shown. The magnetic order is represented by short and long arrows
  that represent small respectively large magnetic moments of the two
  symmetry-inequivalent Ni cations.
\label{fig:nickelate-layer}}
\end{figure}

The strong coupling between electronic and structural degrees of
freedom in the rare-earth nickelates is apparent from the observation
that the MIT is accompanied by a structural transition from $Pbnm$ to
$P2_1/n$. Hereby, formerly symmetry-equivalent NiO$_6$ octahedra
become inequivalent. One half of the NiO$_6$ octahedra expand their
volumes while the other half reduce their volumes by changing the Ni-O
bond lengths accordingly. This results in a three-dimensional
checkerboard-like arrangement of alternating long bond (LB) and short
bond (SB) octahedra.~\cite{Alonso:2000fz,Alonso:2001bs} The $P2_1/n$
structure of LuNiO$_3$ below the MIT~\cite{Alonso:2001bs} is depicted
in Fig.~\ref{fig:exp-struc}. The checkerboard-like arrangement of LB
and SB octahedra around the Ni cations within a
$\left[001\right]$-type plane (in pseudocubic notation) is also
schematically shown in Fig.~\ref{fig:nickelate-layer}.

The exact mechanism that drives this unusual MIT is still under
debate. In the simplest picture, the nominal Ni$^{3+}$ cations are
split into Ni$^{2+}$ and Ni$^{4+}$, corresponding to LB and SB
octahedra, respectively. Such charge disproportionation has been
suggested as alternative to a Jahn-Teller distortion for cases where
the Hund's coupling $J$ overcomes the on-site Coulomb repulsion
$U$.~\cite{Mazin:2007jx} This can occur in systems such as the
nickelates, with a small or negative charge transfer energy and strong
hybridization between Ni $e_g$ orbitals and O $p$ states, resulting in
strong screening of the local Coulomb repulsion and thus a rather
small effective $U$.~\cite{Subedi:2015en} Indeed, no Jahn-Teller
distortion has been observed in the nickelates,~\cite{Zhou:2004jb}
even though the nominal $t_{2g}^6 e_g^1$ electron configuration of the
Ni$^{3+}$ cations should in principle be susceptible to this type of
distortion.
However, it has also been questioned whether a picture of charge
disproportionation on the Ni sites is really
adequate.~\cite{Park:2012hg,Johnston:2014ca} Instead, is was shown,
that a description in terms of ligand holes, delocalized over the O
octahedra, can account for the observed bond-disproportionation
without the need for charge transfer between the Ni
sites.~\cite{Park:2012hg,Johnston:2014ca}

Very recently, \citeauthor{Varignon:2017is} suggested that both
pictures could be consolidated through the use of Wannier functions
centered at the Ni sites, and thus representing the formal valence
states of the Ni cations, but also with significant orbital weight on
the surrounding O ligands.~\cite{Varignon:2017is} Indeed, the minimal
low energy description employed by \citeauthor{Subedi:2015en} is based
on such Ni-centered $e_g$ Wannier functions that are spatially more
extended than simple atomic orbitals.~\cite{Subedi:2015en} We note
that all proposed mechanisms have in common that they result in a
strong modulation of the magnetic moment on the two inequivalent Ni
sites, in the limiting case with a spin $S=1$ on the LB site and $S=0$
on the SB site.

Apart from the exact mechanism underlying the MIT, the magnetic order
observed in the nickelates is also not yet fully resolved and poses
numerous open questions. All systems from $R$=Lu to Pr exhibit the
same antiferromagnetic wave-vector $k= \left[ \frac{1}{4},
  \frac{1}{4}, \frac{1}{4} \right] \cdot \frac{\pi}{a_\text{c}}$
relative to the underlying simple cubic perovskite structure (with
approximate cubic lattice constant
$a_\text{c}$).~\cite{GarciaMunoz:1992if,GarciaMunoz:1994dk,Garcia-Munoz_Rodriguez-Carvajal_Lacorre:1992}
Furthermore, it is known from experiment that the magnetic moments
vary between the LB and SB Ni
sites.\cite{FernandezDiaz:2001ir,Munoz:2009go} However, the exact
magnetic structure is not yet established, due to the lack of
sufficiently large single crystals. There are several possible
arrangements that cannot be distinguished within the experimental
resolution. As a result, it is still under debate whether the magnetic
order is collinear or not.
Moreover, below $\sim$10~K, the magnetic moments of the rare-earth
ions also order. However, while Ref.~\onlinecite{FernandezDiaz:2001ir}
reports a different magnetic periodicity of the rare earth moments
relative to the Ni moments in HoNiO$_3$,
Ref.~\onlinecite{Munoz:2009go} suggests the same periodicity of Ni and
Dy moments in DyNiO$_3$.

In order to gain further insights into the underlying mechanisms, and
also to enable quantitative predictions about the physical properties
of rare-earth nickelates, a first principles-based computational
approach is very desirable. However, an accurate quantitative
description of the complex interplay between the various factors that
are believed to control the MIT in these materials, i.e., structural
properties, electronic correlation effects, and hybridization between
the Ni $3d$ states and the surrounding O ligands, is rather
challenging.

Several previous studies have reported that both structural as well as
electronic and magnetic properties of rare-earth nickelates can, at
least to a certain extent, be described within the ``plus Hubbard
$U$'' extension of density functional theory
(DFT).~\cite{Giovannetti:2009ba,Prosandeev:2012fo,Park:2012hg,He:2015ey,Varignon:2017is}
Specifically, the correct bond-disproportionated crystal structure as
well as the complex AFM order, compatible with experimental
observations, can be obtained from such calculations. On the other
hand, the small energy differences between the different possible
magnetic structures are very difficult to
resolve,~\cite{Giovannetti:2009ba} and in most cases the ferromagnetic
(FM) configuration appears to be energetically more
favorable.~\cite{Prosandeev:2012fo,Park:2012hg} Moreover, it has
recently been stated that DFT+$U$ overestimates the tendency for
bond-disproportionation, and that the more sophisticated DFT plus
dynamical mean-field theory (DMFT) (see, e.g.,
Ref.~\onlinecite{Held:2007fi}) is required for a more accurate
description.~\cite{Park:2014cv}

However, all of these previous studies were either focused on only one
(or few) specific member(s) of the series or have used specific values
for the Hubbard interaction parameter $U$, ranging from $U=2$\,eV to
$U=8$\,eV. It is therefore difficult to draw general conclusions
regarding the predictive capabilities of the DFT+$U$ approach for the
whole series of rare-earth nickelates. In a very recent study,
\citeauthor{Varignon:2017is} have shown that for a small value of
$U=2$\,eV the complex AFM configurations become energetically more
favorable than FM order and that simultaneously good structural
parameters are obtained for a large part of the
series.~\cite{Varignon:2017is} However,
Ref.~\onlinecite{Varignon:2017is} has excluded the members of the
series with the smallest rare earth cations ($R$=Lu to Ho), and the
$U$ dependence of the structural, electronic, and magnetic properties,
as well as the effect of magnetic order on the bond disproportionation
has not been discussed.

Here, we present a systematic study of the whole nickelate series with
rare-earth ions Lu, Er, Ho, Y, Dy, Gd, Sm, Nd, Pr and La, using the
DFT+$U$ formalism. In particular we address the interplay between the
strength of the Hubbard $U$, different magnetic orders, the size of
the rare-earth cation $R$, and the resulting structural
parameters. Our results thus fill an important gap and further clarify
the capabilities of the DFT+$U$ approach for quantitative predictions
of the physical properties of rare-earth nickelates. Our work can also
serve as starting point for further studies using more advanced
electronic structure methods such as, e.g., DFT+DMFT.

Our findings show that the amplitude of the bond-disproportionation
distortion is strongly influenced by the size of $U$ and also by the
specific magnetic order, but that in principle good agreement with
available experimental data can be obtained for $U \approx
2$\,eV. Moreover, the trends across the series agree well with
experiment. Furthermore, our calculations show that a magnetic order
with the experimentally observed wave-vector is energetically favored
for relatively small $U$ values (also around $U \approx 2$\,eV), and
that the energy gain relative to ferromagnetic order increases from Lu
to Pr, consistent with the observed trend of the magnetic ordering
temperature.

The remainder of this article is organized as follows. First, in
Sec.~\ref{sec:mode-decomp}, we introduce the symmetry-based
decomposition of distortion modes and discuss its application to the
experimental structure of LuNiO$_3$.~\cite{Alonso:2001bs} In
Sec.~\ref{sec:theory} we then briefly describe our computational setup
and list all relevant parameters used in the DFT+$U$ calculations. The
presentation of our main results is divided into two parts. We start
in Sec.~\ref{sec:results_exp} by discussing calculations for LuNiO$_3$
based on the experimental structure taken from
Ref.~\onlinecite{Alonso:2001bs}. Hereby, we investigate the stability
of different magnetic phases for different interaction parameters $U$
and $J$ without relaxing the structural degrees of freedom. We then
start to incorporate structural effects by varying the amplitude of
the breathing mode distortion while keeping all other structural
parameters fixed to the experimental values. Finally, our results of
the full structural relaxations across the nickelate series are
presented in Sec.~\ref{sec:results_rel}, and in Sec.~\ref{sec:summary}
we summarize our main results and discuss their implications.

\section{Description of structural distortions using symmetry-based mode decomposition}
\label{sec:mode-decomp}

\begin{table*}[th]
  \centering
    \caption{Mode decomposition of several structures obtained for
      LuNiO$_3$. The top three lines are based on available
      experimental data corresponding to different temperatures:
      $T=643$~K ($Pbnm$)~\cite{Alonso:2001bs}, $T=533$~K
      ($P2_1/n$)~\cite{Alonso:2001bs}, and $T=295$~K
      ($P2_1/n$)~\cite{Alonso:2000fz}. The bottom five lines
      correspond to results of our structural relaxations for
      different magnetic orders (NM: nonmagnetic, FM: ferromagnetic,
      $T$-AFM: $T$-type antiferromagnetic) and using different values
      for the parameters $U$ and $J$.}
\label{tab:dist-comp}
  \begin{ruledtabular}
\begin{tabular}{l c c c c c c c c}
\vspace{0.05cm}
& \multicolumn{5}{c}{$Pbnm$ modes} & \multicolumn{3}{c}{additional $P2_1/n$ modes} \\
\cline{2-6} \cline{7-9} 
& $R_4^+$ & $R_5^+$ & $X_5^+$ & $M_2^+$ & $M_3^+$ & $R_1^+$ & $R_3^+$ & $M_5^+$ \\ \hline
exp. $643$ K ($Pbnm$)   & 0.811 & 0.117 & 0.449 & 0.018 & 0.626 & $-$ & $-$ & $-$ \\ 
exp. $533$ K ($P2_1/n$) & 0.821 & 0.124 & 0.452 & 0.025 & 0.617 & 0.077 & 0.001 & 0.013 \\ 
exp. $295$ K ($P2_1/n$) & 0.826 & 0.124 & 0.454 & 0.031 & 0.616 & 0.077 & 0.002 & 0.007 \\ \hline
NM $U=0$, $J=0$~eV & 0.845 & 0.129 & 0.487 & 0.023 & 0.625 & $-$ & $-$ & $-$ \\ 
FM $U=5$, $J=1$~eV & 0.875 & 0.126 & 0.476 & 0.027 & 0.623 & 0.094 & 0.006 & 0.023 \\ 
$T$-AFM $U=0$, $J=0$~eV & 0.861 & 0.130 & 0.480 & 0.026 & 0.622 & 0.037 & 0.001 & 0.002 \\ 
$T$-AFM $U=2$, $J=1$~eV & 0.870 & 0.129 & 0.480 & 0.034 & 0.625 & 0.081 & 0.008 & 0.019 \\ 
$T$-AFM $U=5$, $J=1$~eV & 0.879 & 0.124 & 0.475 & 0.037 & 0.623 & 0.124 & 0.009 & 0.029 \\ 
\end{tabular}
\end{ruledtabular}
\end{table*}

For a systematic and quantitative discussion of the various structural
distortions that are present in the $Pbnm$ and $P2_1/n$ crystal
structures of the rare-earth nickelates, we use a symmetry-based mode
decomposition as described by
\citeauthor{PerezMato:2010ix}.~\cite{PerezMato:2010ix} Thereby, the
atomic positions within a distorted crystal structure (low-symmetry
structure), $\vec{r}_i^{\ \text{dist}}$, are written in terms of the
positions in a corresponding non-distorted reference structure
(high-symmetry structure), $\vec{r}_i^{\ 0}$, plus a certain number of
independent distortions described by orthonormal displacement
vectors, $\vec{d}_{im}$, and corresponding amplitudes, $A_{m}$:
\begin{align}
    \vec{r}_i^{\ \text{dist}} = \vec{r}_i^{\ 0} + \sum\limits_m A_m \ \vec{d}_{im} \qquad .
\end{align}

The amplitudes $A_m$ can thus be viewed as distinct order parameters
for the different structural distortions present in the low symmetry
structure. This allows to clearly identify the most relevant
structural degrees of freedom, and, in particular for the case of the
rare-earth nickelates, to systematically distinguish between the
various octahedral tilt distortions and the breathing mode related to
the MIT.

The mode displacement vectors $\vec{d}_{im}$ are constructed such that
each mode $m$ has a well-defined symmetry, i.e., it corresponds to a
specific irreducible representation (irrep) of the high symmetry space
group. Here, we use the ideal cubic perovskite structure as high
symmetry reference structure. Thus, all distortion modes are labeled
according to the irreps of space-group $Pm\overline{3}m$. Note that an
irrep can involve distortion vectors with multiple degrees of freedom,
e.g., corresponding to displacement patterns of different inequivalent
atoms. All distortion modes corresponding to the same irrep can then
be grouped together to define a total mode amplitude of that symmetry.

\citeauthor{Balachandran:2013cg} have presented such a symmetry-based
mode decomposition for the low temperature $P2_1/n$ structure of
various nickelates, based on available experimental
data.~\cite{Balachandran:2013cg} Eight different irreps of the high
symmetry $Pm\bar{3}m$ space group can occur within $P2_1/n$. Five of
them, corresponding to symmetry labels $R_4^+$, $M_3^+$, $X_5^+$,
$R_5^+$, and $M_2^+$, are already allowed within the high-temperature
$Pbnm$ structure. The first two of these, $R_4^+$ and $M_3^+$,
correspond to out-of-phase and in-phase tilts of the oxygen octahedra,
and, as shown by
\citeauthor{Balachandran:2013cg},~\cite{Balachandran:2013cg} only the
distortions corresponding to $R_4^+$, $M_3^+$, and $X_5^+$ have
consistently non-negligible mode amplitudes throughout the nickelate
series. In contrast, the amplitude of the $M_2^+$ mode, which
corresponds to a staggered Jahn-Teller distortion of the oxygen
octahedra that is found, e.g., in many manganites, is negligibly
small.

\begin{figure}[t]
  \centering \includegraphics[width=1.0\columnwidth]{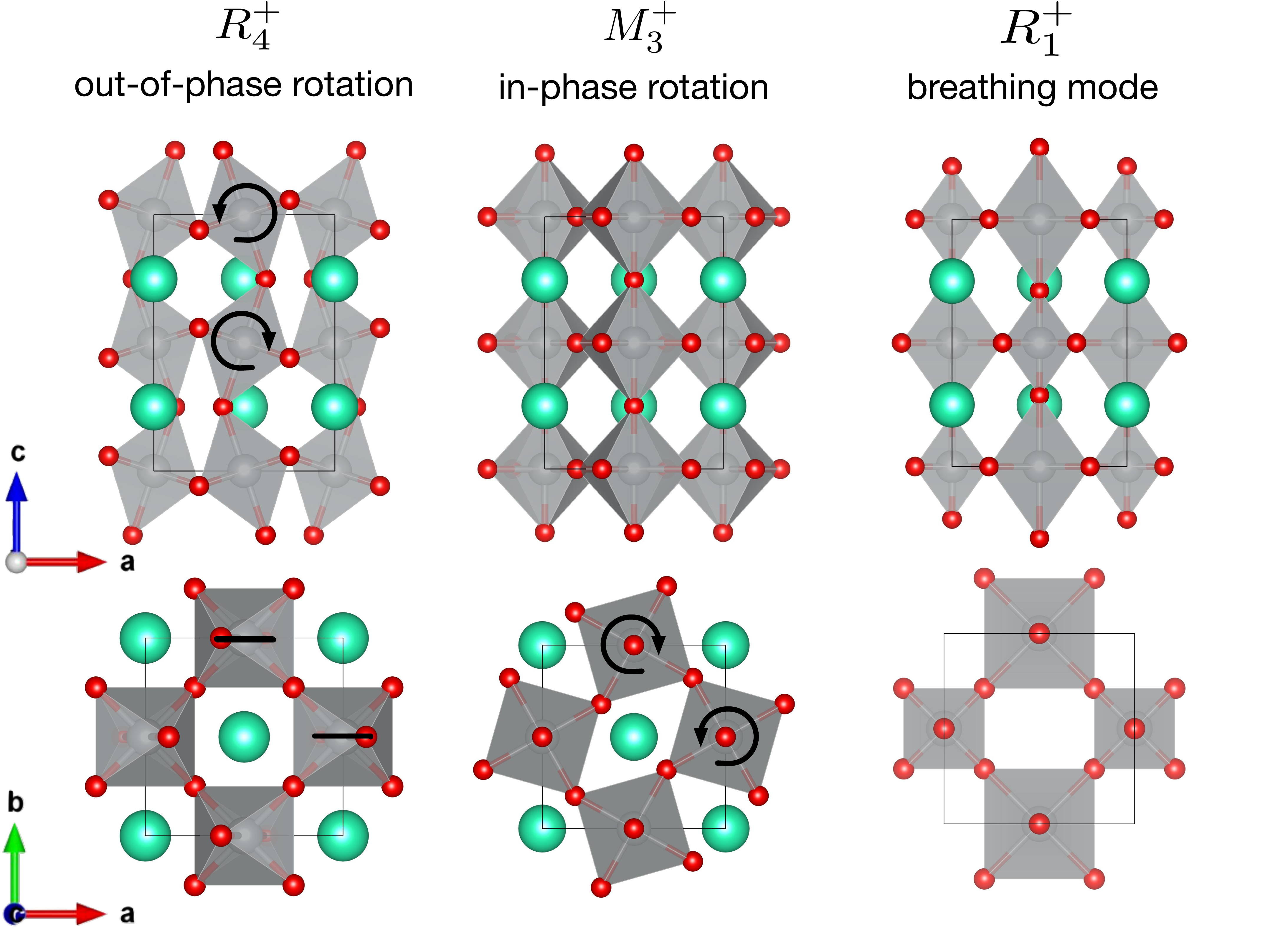}
\caption{Depiction of the three most important distortion modes found
  in nickelate compounds. NiO$_6$ octahedra are shown in gray. The
  $R_4^+$ $Pbnm$ mode that corresponds to an out-of-phase rotation of
  NiO$_6$ octahedra around the orthorhombic $b$ axis, the $M_3^+$
  $Pbnm$ mode that corresponds to an in-phase rotation of NiO$_6$
  octahedra around the $c$ axis, and the $R_1^+$ $P2_1/n$ mode that
  corresponds exactly to the bond-disproportionation of NiO$_6$
  octahedra in the low temperature phase of the nickelates.}
\label{fig:dist-comp}
\end{figure}

The low-temperature $P2_1/n$ structure allows for three additional irreps,
labeled $R_1^+$, $R_3^+$, and $M_5^+$. Here, only the $R_1^+$ mode, which
describes the breathing mode distortion with alternating LB and SB octahedra,
has a non-negligible amplitude.~\footnote{This is true for all nickelates
analyzed in Ref.~\onlinecite{Balachandran:2013cg} except for NdNiO$_3$, which
exhibits a surprisingly large $M_5^+$ mode amplitude. This mode describes an
out-of-phase tilting of oxygen octahedra. However, the relatively large
$M_5^+$ amplitude found in NdNiO$_3$ could be an experimental artifact.} The
three most relevant distortion modes found in the rare-earth nickelates, i.e.,
the octahedral tilt modes $R_4^+$ and $M_3^+$ ($Pbnm$ symmetry), as well as
the $R_1^+$ breathing mode (within $P2_1/n$) are visualized in Fig.~\ref{fig:dist-comp}.

In Table~\ref{tab:dist-comp}, we list the distortion mode amplitudes
for LuNiO$_3$ at the three different temperatures measured in
Refs.~\onlinecite{Alonso:2000fz} and \onlinecite{Alonso:2001bs}. We
use ISODISTORT \cite{Campbell:2006eja} for the calculation of the
distortion mode amplitudes. All mode amplitudes are given in \r{A} and
are normalized to the cubic high-symmetry parent structure (not to the
20 atom $Pbnm$ unit cell). Note that the data at room temperature is
identical (except for the different normalization) to the
corresponding data in the paper of
\citeauthor{Balachandran:2013cg}.~\cite{Balachandran:2013cg}
Table~\ref{tab:dist-comp} also contains data from our structural
relaxations which will be discussed in Sec.~\ref{sec:results_rel}.

If one compares the experimental data from
Ref.~\onlinecite{Alonso:2001bs} obtained approximately 60~K above and
below the MIT (first and second row of Table~\ref{tab:dist-comp}), one
can see that the largest $Pbnm$ mode amplitude, i.e., $R_4^+$, does
almost not change during the MIT. The $R_4^+$ amplitude is
$0.811$\,\r{A} within the high temperature $Pbnm$ phase and
$0.821$\,\r{A} within the low-temperature $P2_1/n$ phase. This value
corresponds to maximal displacements of individual oxygen atoms by
$0.58$\,\r{A}. Similar behavior can be observed for the other two main
modes, $M_3^+$ and $X_5^+$. Finally, the bond-disproportionation mode
in the low-temperature phase, $R_1^+$, exhibits an amplitude of
$0.077$ \r{A}, which corresponds to a displacement of each oxygen atom
by $0.044$ \r{A}.

\section{Computational method}
\label{sec:theory}

DFT calculations are performed using the projector augmented wave
(PAW) method~\cite{Blochl:1994dx} implemented in the ``Vienna Ab
initio Simulation Package''
(VASP).~\cite{Kresse:1993bz,Kresse:1996kl,Kresse:1999dk} We use the
generalized gradient approximation (GGA) in the form proposed by
Perdew, Burke, and Ernzerhof (PBE) \cite{Perdew:1996iq} as exchange
correlation functional. For an improved treatment of the strong local
electron-electron interaction between the Ni 3$d$ electrons, we add an
effective on-site Coulomb interaction $U$ and Hund's rule exchange
interaction $J$ in the form discussed by
\citeauthor{Liechtenstein:1995ip}~\cite{Liechtenstein:1995ip} The
values for $U$ and $J$ are varied throughout this paper as described in
Secs.~\ref{sec:results_exp} and ~\ref{sec:results_rel}. 

For Ni, the 3$p$ semicore states are included as valence electrons in
the PAW potential. For the rare-earth atoms, we use PAW potentials
corresponding to a $3+$ valence state with $f$-electrons frozen into
the core and, depending on the rare-earth cation, the corresponding
$5p$ and $5s$ states are also included as valence electrons. Thus, we
neglect the ordering of the rare-earth $f$ magnetic moments, which
only occurs at very low
temperatures.~\cite{FernandezDiaz:2001ir,Munoz:2009go} The kinetic
energy cut-off for the plane-wave basis is set to 550~eV.

\begin{figure}[t]
\centering \includegraphics[width=1.0\columnwidth]{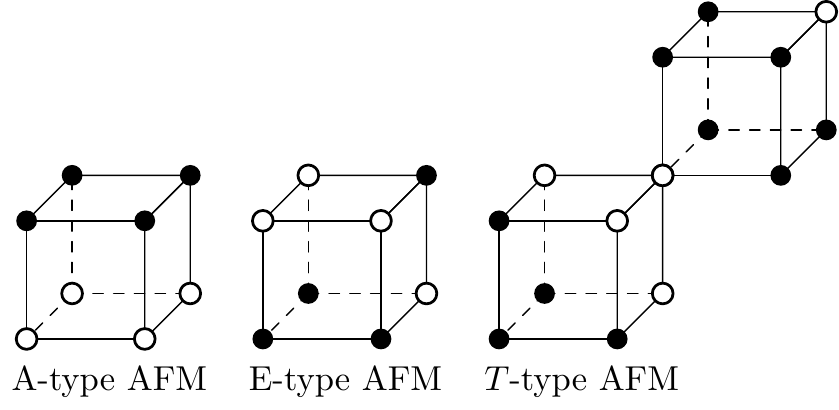}
\caption{Schematic depiction of the 3 different antiferromagnetic
  orderings that are investigated in this paper. White spheres
  correspond to spin-up and black spheres to spin-down moments on Ni
  sites. The special ordering along the [111] pseudocubic crystal
  axis is labeled $T$-AFM\cite{Giovannetti:2009ba}. For the sake of
  simplicity we dropped all rare-earth atoms and oxygen atoms.}
\label{fig:magnet-orderings}
\end{figure}

We consider four different types of magnetic order: FM and three
different types of AFM order, which are depicted in
Fig.~\ref{fig:magnet-orderings}. The rather common $A$-type AFM order
($A$-AFM) corresponds to an alternating ($\uparrow \downarrow$)
alignment of Ni magnetic moments along the Cartesian $z$ direction,
and parallel alignment perpendicular to $z$. The $E$-type AFM order
($E$-AFM) corresponds to an $\uparrow \uparrow \downarrow \downarrow$
alignment along both $x$ and $y$ directions, and alternating moments,
i.e. $\uparrow \downarrow \uparrow \downarrow$, along the $z$
direction. The $T$-type AFM order, which corresponds to the
experimentally observed AFM wave vector, $k= \left[ \frac{1}{4},
  \frac{1}{4}, \frac{1}{4} \right] \cdot \frac{\pi}{a_\text{c}}$,
exhibits an $\uparrow \uparrow \downarrow \downarrow$ pattern along
all three Cartesian directions.
As shown by \citeauthor{Giovannetti:2009ba} \cite{Giovannetti:2009ba}
using DFT+$U$ calculations, this $T$-type AFM order is energetically
nearly indistinguishable from the other two magnetic order patterns
(one collinear and one noncollinear) that are compatible with the
experimental neutron data. In the following, we therefore use the
(relatively simple) $T$-AFM structure as representative for the
experimentally observed magnetic order.
Both $T$-AFM and bond-disproportionation are also illustrated in
Fig.~\ref{fig:nickelate-layer} within a [001]-type layer.

For the examination of the different magnetic order patters, different
unit cell sizes are used. For the FM and $A$-AFM order, we use a 20
atom unit cell consisting of $\sqrt{2} \times \sqrt{2} \times 2$
(pseudo-) cubic perovskite units. This cell also corresponds to the
primitive crystallographic unit cells for both the $Pbnm$ and $P2_1/n$
structures. For the $E$-AFM magnetic structure, this unit cell is
doubled along the $a$ direction (40 atoms), and for the special
$T$-AFM order the cell is doubled once more, this time along the $c$
direction (80 atoms). 
A $k$-point mesh with $10 \times 10 \times 8$ grid points along the
three reciprocal lattice directions is used for the 20 atom $Pbnm$ and
$P2_1/n$ unit cells to perform Brillouin zone integrations. For the 40
atom $E$-AFM cell we use an appropriately reduced $5 \times 10 \times
8$ $k$-point grid and for the 80 atom $T$-AFM cell a $5 \times 10
\times 4$ grid. For accurate structural relaxations, the forces acting
on all atoms are minimized until all force components are smaller than
$10^{-4}$ eV/\r{A}. Local magnetic moments are obtained by integrating
the spin density within the PAW spheres (LORBIT=11).

\section{Calculations for LuNiO$_3$ based on the experimentally observed structure}
\label{sec:results_exp}

\begin{figure}[t]
\centering \includegraphics[width=1.0\columnwidth]{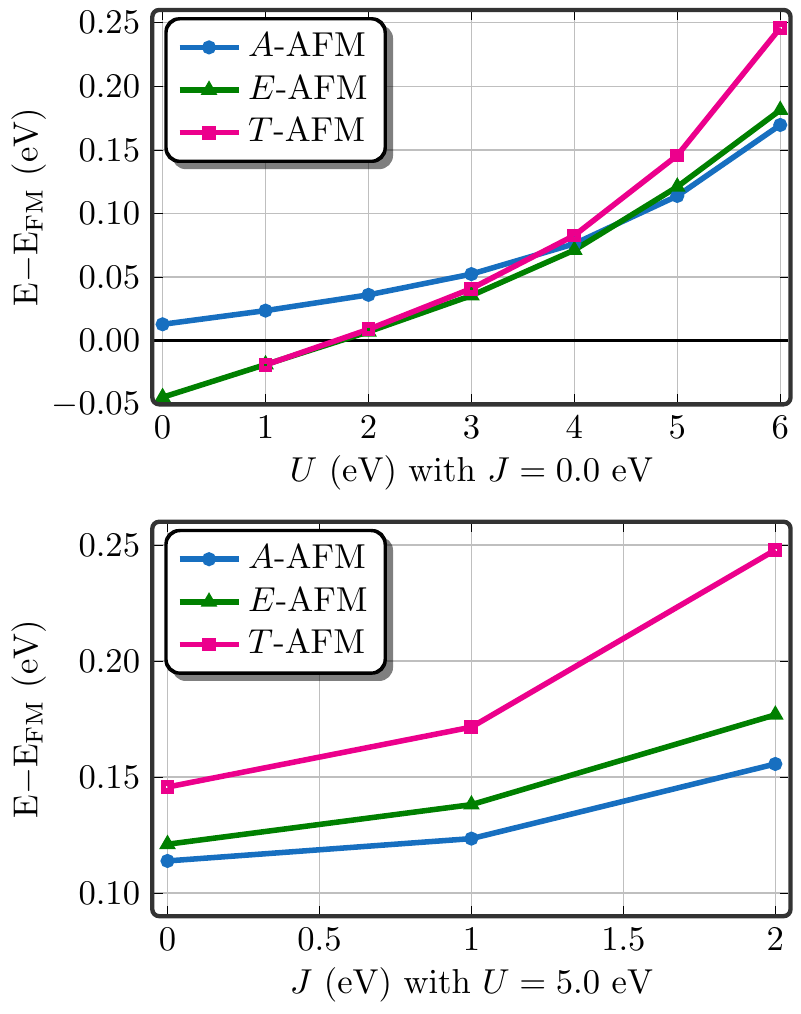}
\caption{Effect of $U$ (top) and $J$ (bottom) on the energy
  differences between different magnetically ordered states,
  calculated for LuNiO$_3$ using the experimentally observed $P2_1/n$
  structure. For each $U$ and $J$, the energies of three different AFM
  states are given relative to the FM state, normalized to a unit cell
  containing 4 Ni sites. The $T$-AFM order (magenta squares) is
  energetically favored only for rather small $U$ values. Increasing
  $J$ (bottom) also favors the FM state.}
\label{fig:lunio3-exp-u-comp}
\end{figure}


We start by performing calculations for LuNiO$_3$ in the
experimentally determined $P2_1/n$ structure at
\mbox{$T=533$~K},~\cite{Alonso:2001bs} and analyze the influence of
the Hubbard and Hund's rule interaction parameters, $U$ and $J$, on
the relative stability of different magnetic configurations. We
consider the FM case as well as three different AFM configurations
($A$-AFM, $E$-AFM, and $T$-AFM, see Sec.~\ref{sec:theory} for more
details). Fig.~\ref{fig:lunio3-exp-u-comp} shows the calculated total
energies of the three AFM configurations relative to the FM state. All
energies are normalized to a 20 atom unit cell. A negative value
indicates that the corresponding AFM state is lower in energy, whereas
a positive value indicates that the FM state is lower in energy.

The top panel of Fig.~\ref{fig:lunio3-exp-u-comp} depicts the case
with $J=0$.  It can be seen that for small values of $U$, the $T$-AFM
and $E$-AFM states have very similar energies, and for $U<2$\,eV, both
are lower in energy than the FM state. For values $U>2$\,eV the FM
state is most favorable, while the simple $A$-AFM state is higher in
energy over the whole range of $U$ values. We also note that for $U=0$
we were not able to stabilize the $T$-AFM state within our
calculations.

The bottom panel of Fig.~\ref{fig:lunio3-exp-u-comp} shows the effect
of varying the Hund's coupling parameter $J$ for fixed $U=5$\,eV.
Consistent with the results shown in the top panel, at this $U$ value
the FM state is favored for $J=0$. If $J$ is increased, the FM state
becomes even more favorable compared to all three AFM orderings. The
same trend can be observed for other values of $U$ (not
shown). Increasing $J$ lowers the energy of the FM state relative to
the various AFM orderings.

It appears that the $T$-AFM state, i.e., the state that is compatible
with the experimental observations, is only favorable for small values
of $U$ and $J$. Furthermore, within this range of $U$ and $J$, the
energy difference between $T$-AFM and the closely related $E$-AFM state
is rather small. On increasing both $U$ and $J$, the FM states becomes
lower in energy than all considered AFM orderings.

We note that nonmagnetic DFT calculations with $U=0$ for LuNiO$_3$ in
both the low temperature $P2_1/n$ and the high temperature $Pbnm$
structures (taken from Ref.~\onlinecite{Alonso:2001bs}) result in a
metallic system. By adding the local Coulomb interaction $U$ we are
able to stabilize the $T$-AFM order, which then results in an
insulating ground state. FM order also results in an insulating ground
state for $U>0$\,eV. Thus, magnetic order and a small value of $U$
(around 1\,eV or larger) is needed to obtain an insulating ground
state in the experimental low temperature structure.

\begin{figure*}[t]
\centering
\includegraphics[width=1.0\columnwidth]{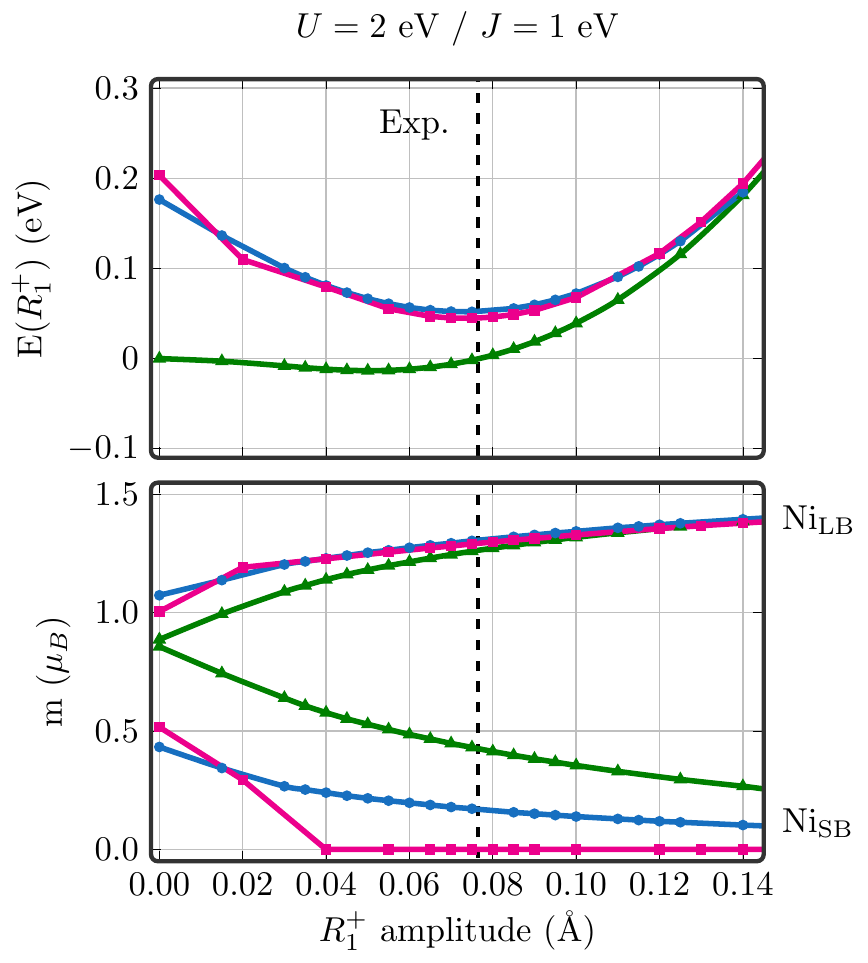}
\hfill
\includegraphics[width=1.0\columnwidth]{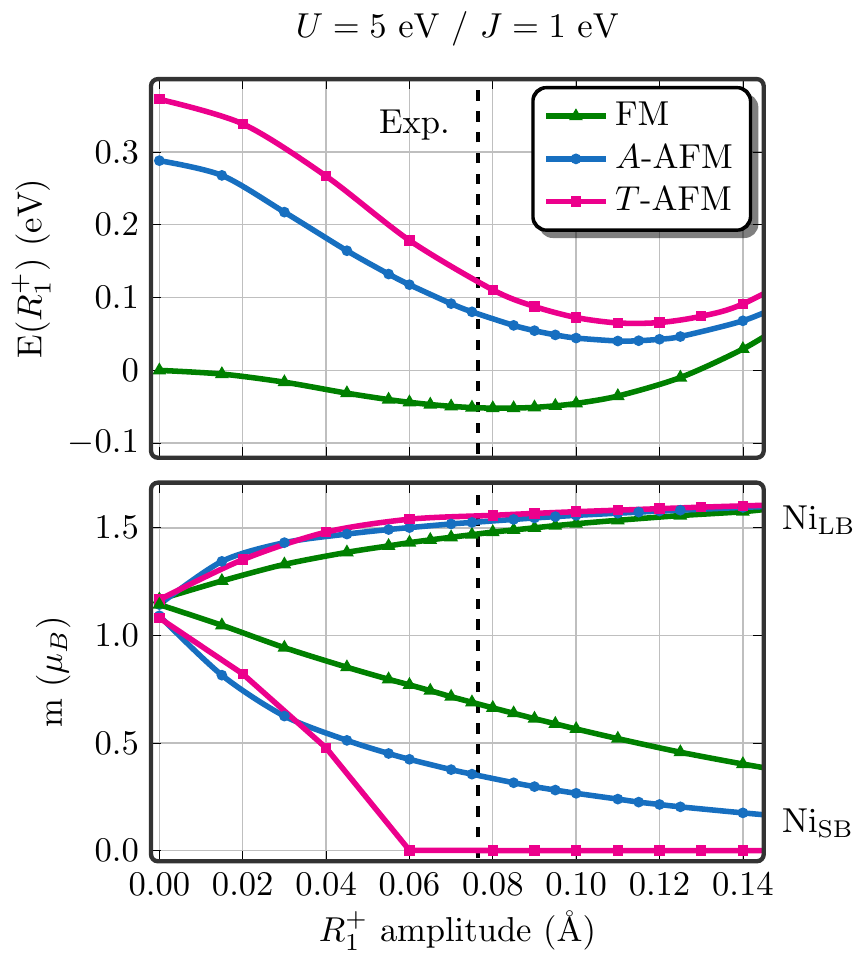}
\caption{Total energy (top) and Ni magnetic moments (bottom) as
  function of the $R_1^+$ mode amplitude for different magnetic states
  and $U$ values (left: $U=2$\,eV, right: $U=5$ eV) and $J=1$\,
  eV. All other structural parameters (apart from the $R_1^+$ mode)
  are fixed to the experimental $P2_1/n$ structure, and the $R_1^+$
  mode amplitude in the experimental structure is marked by the
  vertical dashed lines. In the bottom panels, the upper (lower)
  curves correspond to the Ni$_\text{LB}$ (Ni$_\text{SB}$) site.}
\label{fig:lunio3-exp-R1+-mode}
\end{figure*}

To investigate how sensitive the energy differences between different
magnetic states depend on small variations in the crystal structure,
in particular the $R_1^+$ breathing mode, we now use the mode
decomposition of the experimental $P2_1/n$ structure and tune the
amplitude of the $R_1^+$ mode while keeping all other structural
degrees of freedom fixed to their experimental values. The result is
shown in Fig.~\ref{fig:lunio3-exp-R1+-mode}, which shows the total
energy (top) and the magnetic moments on the Ni sites (bottom) as
function of the $R_1^+$ mode amplitude for different magnetic
orderings and two different $U$ values ($U=2$\,eV on the left and
$U=5$\,eV on the right) together with $J=1$ eV.
In each case, the energy exhibits a minimum at a finite value of the
$R_1^+$ amplitude, which indicates the value predicted by DFT+$U$ for
a given magnetic order (with all other structural parameters fixed to
experimental values). The black vertical dashed line indicates the
$R_1^+$ amplitude in the experimental structure,
$R_1^+=0.077$\,\r{A}. Note that the experimental structure was
determined in the paramagnetic insulating phase.

One observes that the energy of the FM state (green triangles) is
always lower than that of the AFM states (magenta squares and blue
circles). However, the energy difference between FM and AFM order is
smaller for $U=2$\,eV than for $U=5$\,eV, consistent with the results
shown in Fig.~\ref{fig:lunio3-exp-u-comp} (note that the top panel in
Fig.~\ref{fig:lunio3-exp-u-comp} corresponds to $J=0$ whereas
Fig.~\ref{fig:lunio3-exp-R1+-mode} is obtained using $J=1$\,eV, and
that increasing $J$ favors the FM state). Furthermore, it is apparent
that the AFM states couple much stronger to the $R_1^+$ breathing mode
than the FM state, with a much deeper energy minimum relative to zero
mode amplitude and a position of the energy minimum at significantly
larger $R_1^+$ amplitude.

The predicted mode amplitude for $T$-AFM and $U=2$\,eV ($R_1^+ =0.076$
\r{A}) agrees very well with the experimental value, whereas the FM
state results in an amplitude ($R_1^+=0.050$ \r{A}) that is much
smaller than what is observed experimentally. Increasing $U$ increases
the predicted mode amplitudes for both FM and AFM order, and for
$U=5$\,eV the amplitude obtained for FM order ($R_1^+=0.082$ \r{A}) is
close to the experimental value, whereas both AFM states exhibit
significantly larger amplitudes ($R_1^+ \approx 0.11$ \r{A}). There is
only a small difference between the two different AFM orderings, and
in particular the positions of the energy minima are very similar. For
$U=2$\,eV, $T$-AFM is slightly lower in energy than $A$-AFM, whereas
for $U=5$\,eV, $A$-AFM is lower.

The bottom panels of Fig.~\ref{fig:lunio3-exp-R1+-mode} show the local
magnetic moments of the Ni cations for the different magnetic
orderings as a function of the $R_1^+$ amplitude. For $U=5$\,eV, all
moments have the same value of \mbox{$\sim 1.1\,\mu_\text{B}$} at zero
$R_1^+$ amplitude. In contrast, for $U=2$\,eV, the SB and LB moments
differ already for $R_1^+=0$ in the two AFM cases, while they are both
equal to $\sim\,0.9\,\mu_\text{B}$ in the FM case. It thus appears
that for $U=2$\,eV, the magnetic moments are much more susceptible to
the small symmetry breaking resulting from the presence (albeit with
very small amplitude) of the two other $P2_1/n$ modes, i.e., $R_3^+$
and $M_5^+$. 
We note that if these additional modes as well as the small monoclinic
tilt of the unit cell are also removed, i.e., if the underlying crystal
structure has exact $Pbnm$ symmetry, then the difference between the
LB and SB moments also vanishes in the case of A-AFM order. However,
this is not the case for the $T$-AFM ordering, since $T$-AFM order by
itself breaks the $Pbnm$ symmetry, leading to two
symmetry-inequivalent Ni sites.

With increasing $R_1^+$ amplitude, the moments of the SB sites
decrease and the moments of the LB sites increase. Thereby, the size
of the Ni$_\text{LB}$ moments is rather independent of the magnetic
order, and seems to converge to a value of around
$1.4\,\mu_{\text{B}}$ ($1.6\,\mu_{\text{B}}$) for $U=2$\,eV
($U=5$\,eV). In contrast, the decrease of the Ni$_\text{SB}$ moments
depends more strongly on the magnetic order. For $T$-AFM order, the
Ni$_\text{SB}$ moments vanish completely for $R_1^+$ amplitudes larger
than $R_1^+=0.04$\,\r{A} ($R_1^+=0.06$\,\r{A}) for $U=2$\,eV
($U=5$\,eV). This means that, for both $U$ values, the
Ni$_{\text{SB}}$ moments in the $T$-AFM state are zero at the
experimental $R_1^+$ amplitude. For the $A$-AFM and FM cases, the SB
moments seem to only asymptotically converge to zero, with the
residual moment in the FM case about twice as large as in the $A$-AFM
case.
These results are consistent with earlier studies that also found
nonvanishing magnetic moments on the SB sites for LuNiO$_3$ with FM
order, \cite{Park:2012hg}, and vanishing Ni$_\text{SB}$ moments for
NdNiO$_3$ with $T$-AFM order (for not too large
$U$).\cite{Prosandeev:2012fo}

We note that the behavior of the SB moments for larger $R_1^+$
amplitudes is consistent with a picture where the Ni$_{\text{SB}}$ moment is
simply induced by the effective field created by the magnetic moments
on the neighboring Ni$_{\text{LB}}$ sites. In the FM case, each SB site is
surrounded by six LB nearest neighbors with parallel alignment of
their magnetic moments. In the $A$-AFM case, only four of the six
neighboring LB moments are parallel to each other, and thus the
effective field at the SB site is reduced. For the $T$-AFM case,
exactly half of the neighboring LB moments are aligned parallel to
each other, while the other half is aligned antiparallel, leading to
a cancellation of the effective field on the SB site.
We also performed some calculations where we initiated the magnetic
moments according to $G$-type AFM order. In this case, all LB moments
are parallel to each other and thus the effective field at the SB site
is the same as for FM order. As a result, the calculations converge to
the FM solution even if the SB moments are initiated antiparallel to
the LB moments.

It appears that DFT+$U$ is able to correctly describe the
bond-disproportionated state in LuNiO$_3$, resulting in $R_1^+$
amplitudes that are consistent with the experimentally obtained
structure. However, the precise value of the $R_1^+$ amplitude depends
strongly on the type of magnetic order that is imposed in the
calculation, and also on the value used for the Hubbard interaction parameter $U$.
The complex $T$-AFM state, which is consistent with the available
experimental data and is also stable within the calculations, is lower
in energy than the FM and $A$-AFM states for small values of $U$ (and
$J$). However, the calculations seem to favor the FM solution for $U$
values larger than $U=2$ eV.
Furthermore, the $R_1^+$ breathing mode results in a strong energy
lowering of the AFM states and also leads to a strong
disproportionation between the magnetic moments on the two different
Ni sites (for all magnetic orderings). For $T$-AFM, the local magnetic
moments on the Ni$_\text{SB}$ sites vanish completely at the
experimental $R_1^+$ amplitude.

We note that while different magnetic structures assumed in the
refinements of the available experimental data generally lead to
different values for the local magnetic moments, most studies indeed
report a significant difference between Ni$_\text{LB}$ and
Ni$_\text{SB}$ moments (see, e.g.,
Refs.~\onlinecite{Munoz:2009go,FernandezDiaz:2001ir}). Furthermore,
our $T$-AFM calculations show that the Ni$_\text{SB}$ moments can be
zero, in spite of the fact that the integrated charges inside the PAW
spheres differ only very little between the two different Ni sites
(consistent with previous DFT-based studies). Thus, the SB moments can
vanish completely even though the integrated charges do not correspond
to a naive picture of full charge disproportionation within atomic
spheres.
The results presented in this section are also in good agreement with
a recent DFT+$U$ study by
\citeauthor{Varignon:2017is}\cite{Varignon:2017is} focusing on the
members of the nickelate series with large $R$ cations, which suggests
that a value of $U=2$\,eV gives the best overall agreement with
experimental observations, both regarding magnetic order and the
magnitude of the bond disproportionation.

\section{Structural relaxations for the whole nickelate series}
\label{sec:results_rel}

\begin{figure}[t]
\centering
\includegraphics[width=1.0\columnwidth]{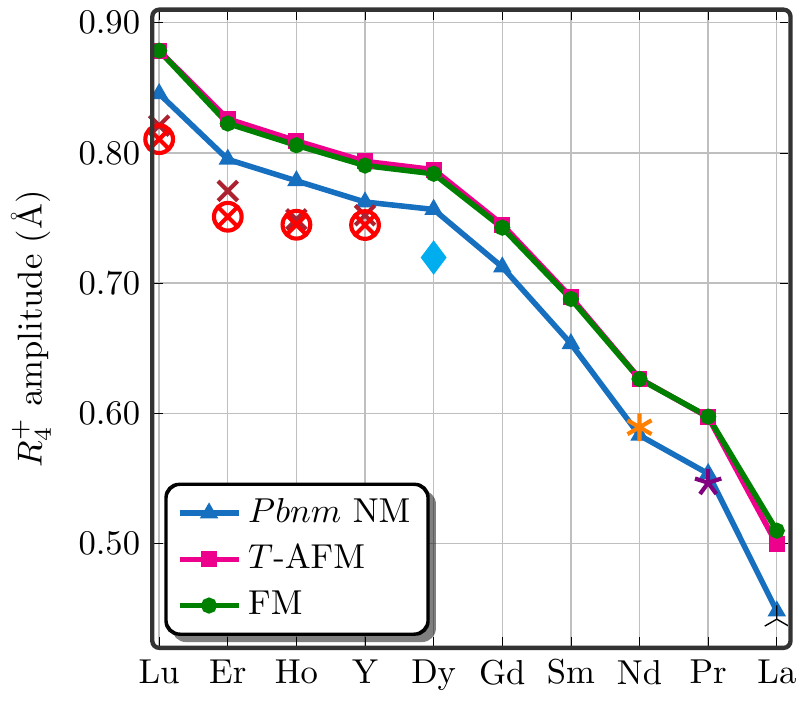}
\caption{$R_4^+$ mode amplitudes obtained from full structural
  relaxations across the nickelate series. The $T$-AFM (magenta
  squares) and the FM calculations (green circles) are performed using
  $U=5$\,eV and $J=1$\,eV, and compared to nonmagnetic (NM)
  calculations (blue triangles) within $Pbnm$ symmetry and using
  $U=0$. The larger symbols not connected by lines indicate various
  experimental results: Lu-Y from Ref.~\onlinecite{Alonso:2001bs}
  (brown crosses obtained at 60\,K below T$_{\text{MIT}}$, red crosses
  with circle at 60\,K above T$_{\text{MIT}}$), Dy (blue diamond) from
  Ref.~\onlinecite{Munoz:2009go}, Nd (orange asterisk) from
  Ref.~\onlinecite{MartinezLope:2009hm}, Pr (purple star) from
  Ref.~\onlinecite{Medarde:2008ge}, and La (black three-pointed star)
  from Ref.~\onlinecite{GarciaMunoz:1992if}.}
\label{fig:r4+-mode}
\end{figure}

Next, we perform full structural relaxations within the
low-temperature $P2_1/n$ symmetry across the whole series of
nickelates with $R$ from Lu to La. We again compare different values
of $U$ and $J$ and different magnetic orderings. However, we will
focus mainly on the FM and $T$-AFM cases, since other AFM orderings
give results similar to $T$-AFM. In addition, we also perform
nonmagnetic (NM) structural relaxations with $U=0$. Note that in this
case the $R_1^+$ breathing mode is not stable and the system relaxes
back to the higher symmetry $Pbnm$ structure, even if we initialize
the system with a finite $R_1^+$ amplitude and $P2_1/n$ symmetry.
To allow for a systematic comparison across the whole series, we also
relax LaNiO$_3$ within both $Pbnm$ and $P2_1/n$ symmetries (i.e.,
similar to all other compounds), even though LaNiO$_3$ is
experimentally found to exhibit a slightly different structure with
$R\bar{3}c$ space group symmetry.~\cite{GarciaMunoz:1992if}

Generally, our calculated lattice parameters agree very well with
available experimental data across the whole series, with maximal
deviations of the unit cell volume of a few percent or less. For
example, for LuNiO$_3$ the NM calculation results in a unit cell
volume that deviates by $-1.5\%$ from the experimental high
temperature structure,~\cite{Alonso:2001bs} whereas the volume
obtained in the FM calculation with $U=5$\,eV and $J=1$\,eV differs by
only $+0.2\%$ from that of the experimental $P2_1/n$ structure at
$\sim$60\,K below $T_\text{MIT}$.~\cite{Alonso:2001bs}

In Table~\ref{tab:dist-comp} we list the amplitudes of all distortion
modes obtained for LuNiO$_3$ in different settings. It can be seen
that the $R_1^+$ mode is the only mode which depends very strongly on
$U$, $J$, and the type of magnetic order. All other relevant mode
amplitudes agree well with the experimental data, except maybe for a
slight overestimation of the $R_4^+$ mode (and perhaps also $X_5^+$),
in particular for the FM/AFM cases and increasing $U$.

As discussed in Sec.~\ref{sec:mode-decomp}, the $R_4^+$ mode is the
most prominent distortion mode in the nickelate series and describes
the out-of-phase octahedral tilts around the in-plane $a$ direction
(Glazer tilt $a^-a^-c^0$, see Fig.~\ref{fig:dist-comp}). The evolution
of the $R_4^+$ amplitude across the nickelate series, calculated for
different settings and compared to experimental data, is depicted in
Fig.~\ref{fig:r4+-mode}. Experimental data for $R$=Lu, Er, Ho, and Y
is taken from the two papers by
\citeauthor{Alonso:1999gk},~\cite{Alonso:1999gk,Alonso:2001bs} for
$R$=Dy from \citeauthor{Munoz:2009go},~\cite{Munoz:2009go} for $R$=Nd
from \citeauthor{MartinezLope:2009hm},~\cite{MartinezLope:2009hm} and
for $R$=Pr from \citeauthor{Medarde:2008ge}.~\cite{Medarde:2008ge}
Note, that the structural data is generally measured at different
temperatures and that
\citeauthor{Alonso:1999gk}~\cite{Alonso:1999gk,Alonso:2001bs} have
obtained data both above and below the MIT transition, i.e., both
within the metallic high temperature $Pbnm$ phase and the insulating
low temperature $P2_1/n$ phase. However, we note that in all these
cases, there is only a rather small difference in the $R_4^+$
amplitude between the two phases (see also Table~\ref{tab:dist-comp}
for the case with $R$=Lu).

The amplitude of the $R_4^+$ mode is monotonously decreasing across
the series from Lu to La, consistent with the increasing radius of the
$R$-cation. Furthermore, the $R_4^+$ amplitude is slightly smaller for
the NM calculation with $U=0$, compared to both FM and $T$-AFM
calculations with $U=5$\,eV and $J=1$\,eV, while there is only a
negligible difference between FM and $T$-AFM. Overall, there is rather
good agreement, both qualitatively and quantitatively, between the
calculated and experimentally measured mode amplitudes. The best
agreement is obtained for the NM case with $U=0$, whereas the magnetic
relaxations with $U=5$\,eV lead to a slight overestimation of the
octahedral tilt distortion compared to the experimental data.

\begin{figure}[t]
\centering
\includegraphics[width=1.0\columnwidth]{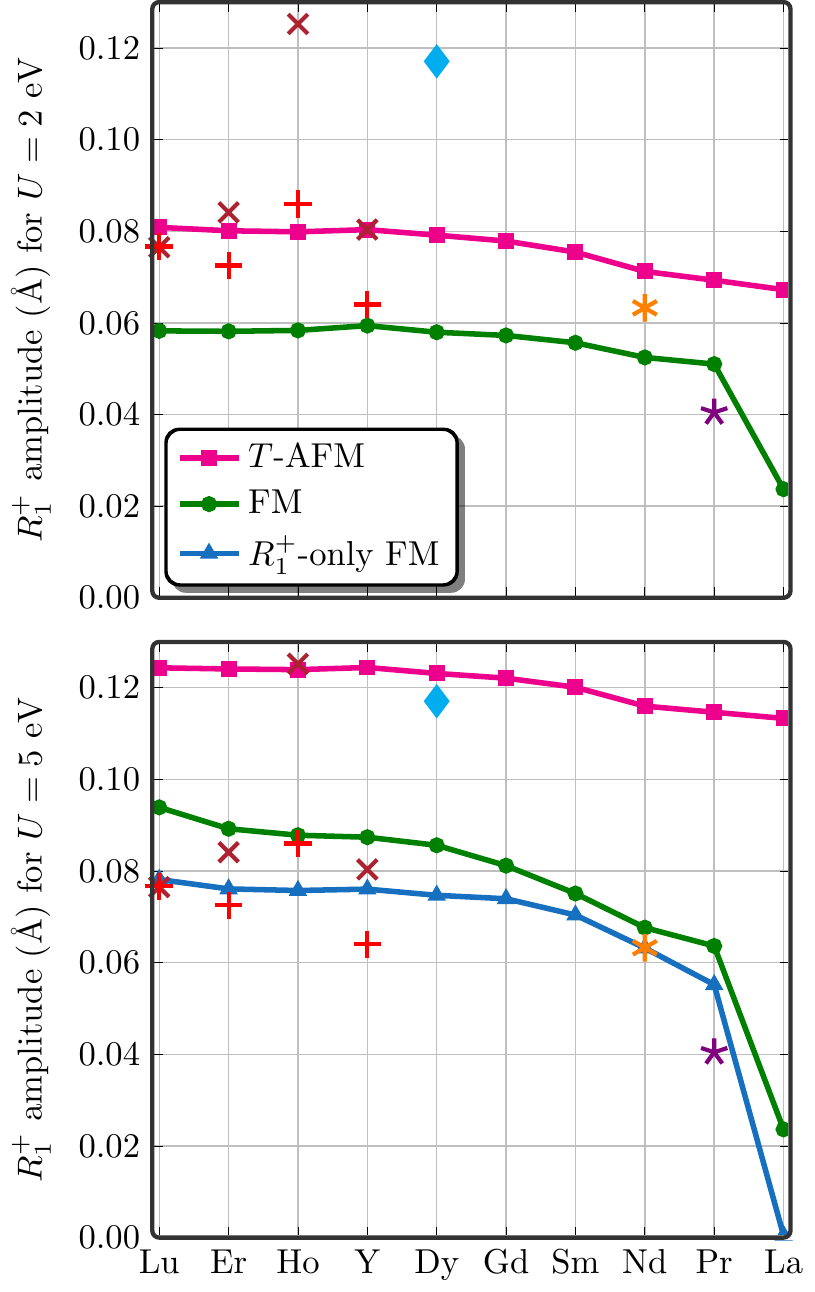}
\caption{$R_1^+$ mode amplitude for the relaxed structures with $U=2$\, eV
  (top) and $U=5$\,eV (bottom), in both cases using $J=1$\,eV. The
  relaxed mode amplitudes are given for the FM (green circles) and
  $T$-AFM (magenta squares) cases, as well as for the ``$R_1^+$-only''
  relaxation with FM order (blue triangles). The large disconnected
  symbols (same in both panels) indicate different experimental
  values: brown crosses for Lu-Y from Ref.~\onlinecite{Alonso:2001bs}
  at 60\,K below $T_{\text{MIT}}$, red plus symbols for Lu-Y from
  Ref.~\onlinecite{Alonso:2000fz} at $T=290$\,K, blue diamond for Dy
  from Ref.~\onlinecite{Munoz:2009go}, orange asterisk for Nd from
  Ref.~\onlinecite{MartinezLope:2009hm}, and purple star for Pr
  from Ref.~\onlinecite{Medarde:2008ge}.}
\label{fig:r1+-mode}
\end{figure}

Next, we discuss the $R_1^+$ breathing mode amplitude. We first note
that, in contrast to the calculations for the fixed experimental
structure presented in the previous section, the $T$-AFM magnetic
order is stable within the fully relaxed structure even in the case
with $U=0$~eV. Moreover, in contrast to the FM and $A$-AFM (and NM)
cases, in the $T$-AFM case all compounds from $R$=Lu to La develop a
finite $R_1^+$ amplitude already for $U=0$. Although the resulting
amplitudes are about two to three times smaller than the
experimentally observed $R_1^+$ amplitudes, this nevertheless
indicates that $T$-AFM strongly supports the $R_1^+$ mode. For larger
$U$ values, a finite $R_1^+$ amplitude emerges from the relaxations
for all considered magnetic orderings.

In the following, we compare results for two different values of $U$,
a smaller value of $U=2$\,eV and a larger value of $U=5$\,eV, in both
cases with $J=1$\,eV. The corresponding $R_1^+$ mode amplitudes for FM
and $T$-AFM cases are shown in Fig.~\ref{fig:r1+-mode} (top:
$U=2$\,eV; bottom: $U=5$\,eV) together with available experimental
data.
Furthermore, to assess whether the slight overestimation of the
$R_4^+$ octahedral tilt mode in the magnetically ordered $+U$
calculations ({\it cf.} Fig.~\ref{fig:r4+-mode}) affect the calculated
$R_1^+$ amplitude, we also consider a third case. Here, we use the
$Pbnm$ structure obtained for the NM case (with $U=0$), and then relax
only the $R_1^+$ amplitude using FM order and $U=5$\,eV (while keeping
all other mode amplitudes fixed). In the following, this relaxation is
referred to as ``$R_1^+$-only''. The corresponding data is also shown
in Fig.~\ref{fig:r1+-mode}.

It can be seen that there are significant differences in the
calculated $R_1^+$ mode amplitudes for the various cases, similar to
what has been found in the previous section for LuNiO$_3$. The
calculated $R_1^+$ mode amplitudes are consistently larger for $T$-AFM
(magenta) compared to the FM case (green), and the larger $U$ value
results in overall larger $R_1^+$ amplitude across the whole
series. Furthermore, in all cases we obtain a decrease of the $R_1^+$
amplitude across the series from $R$=Lu towards $R$=La. This decrease
is most pronounced for the FM case with $U=5$\,eV. The
``$R_1^+$-only'' relaxations (blue) result in reduced $R_1^+$
amplitudes compared to the full FM relaxations at $U=5$\,eV. As
suggested above, this can be attributed to the reduced octahedral tilt
distortion ($R_4^+$ mode) in the underlying NM structures.

Rather good agreement with the experimental data is obtained in the
T-AFM case using $U=2$\,eV, in particular for the compounds at the
beginning of the series. However, the decrease towards $R$=Pr appears
weaker than for the experimental data. For FM order and $U=5$\,eV, the
agreement is also good, including the decrease of the $R_1^+$ towards
the end of the series. Note that a slightly smaller $U$ value would
also slightly reduce the $R_1^+$ amplitude and probably further
improve the comparison of the FM case with the experimental data.

Another fact that becomes apparent from Fig.~\ref{fig:r1+-mode}, is
the rather large scattering of the experimental results for different
members of the series, or even for the same compound measured at
different temperatures (see, e.g., the data for $R$=Ho or Dy in
Fig.~\ref{fig:r1+-mode}). This can be attributed to difficulties in
sample preparation, which is only possible under high pressure,
leading to very small sample sizes and thus low experimental
resolution.~\cite{Medarde:1997vt,Catalan:2008ew} Nevertheless, it
seems that the decrease in $R_1^+$ amplitude for $R$=Nd and in
particular $R$=Pr compared to the compounds at the beginning of the
series can indeed be inferred from the experimental data.

\begin{figure}[t]
\centering
\includegraphics[width=1.0\columnwidth]{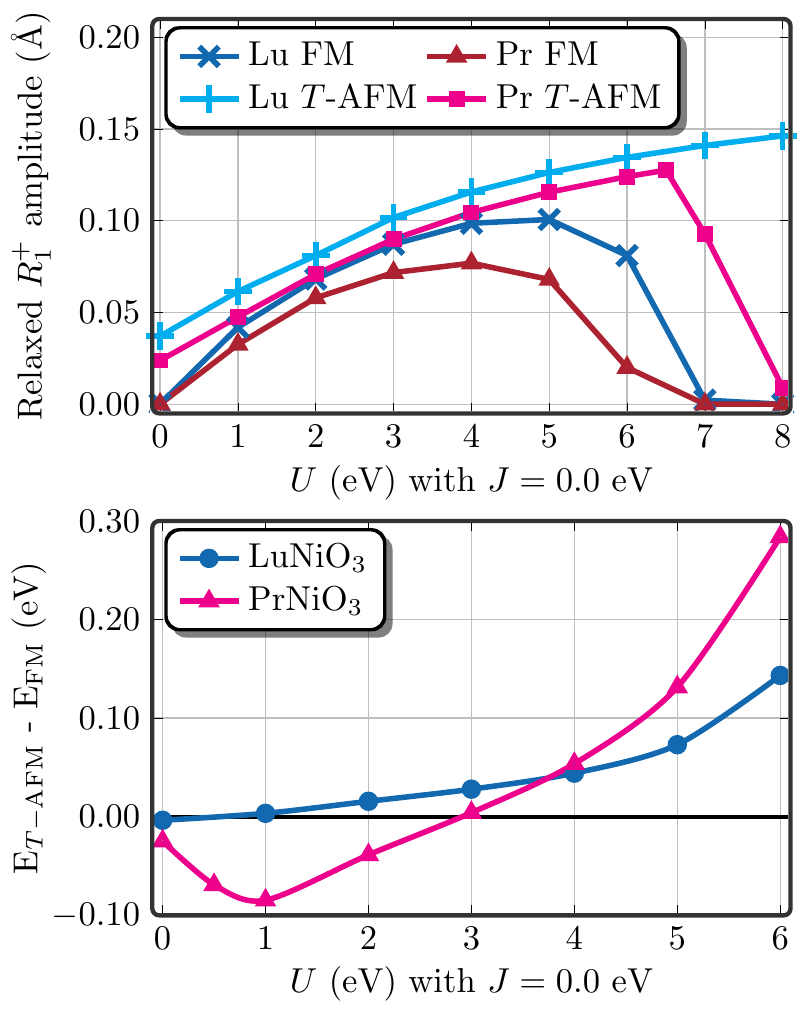}
\caption{Top: $R_1^+$ amplitudes calculated for LuNiO$_3$ (cyan plus
  symbols and blue crosses) and PrNiO$_3$ (brown triangles and magenta
  squares) as a function of $U$ (with $J=0$), in both cases with FM
  (crosses and triangles) as well as $T$-AFM ordering (plus symbols
  and squares). Bottom: energy difference between $T$-AFM and FM
  states for PrNiO$_3$ (magenta triangles) and LuNiO$_3$ (blue
  circles) for different $U$ values (and $J=0$).}
\label{fig:lu-vs-pr-rel}
\end{figure}

We now have a closer look at the $U$ dependence of the $R_1^+$
amplitude across the series. For this, we focus on the two
``end-members'' of the nickelate series, LuNiO$_3$ and PrNiO$_3$, and
perform full structural relaxations for various $U$ values and both FM
and $T$-AFM magnetic orders.  Here, we use $J=0$, so that the limiting
case with $U=0$ can be continuously incorporated. The results are
depicted in the top panel of Fig.~\ref{fig:lu-vs-pr-rel}. We note that
while the $R_1^+$ amplitude is very sensitive to the choice of $U$,
the influence of $J$ is much weaker, and therefore we present only
results for varying $U$.

In agreement with the results shown in Fig.~\ref{fig:r1+-mode}, the
$T$-AFM state leads to an overall larger $R_1^+$ amplitude compared to
the FM state. Furthermore, the $R_1^+$ amplitude is consistently
larger for LuNiO$_3$ than for PrNiO$_3$ (with the same magnetic
order). In all cases, the $R_1^+$ amplitude is monotonously increasing
with $U$ up to about 3-4\,eV. For larger $U$, the $R_1^+$ amplitude
starts to decrease and can even vanish completely at large $U$. The
value of $U$ where the turnaround from increasing to decreasing
$R_1^+$ amplitude occurs, depends both on the $R$ cation and the
magnetic order. It is lowest for Pr and FM order and highest for Lu
and $T$-AFM order (in fact, in this latter case the turnaround does
not occur up to $U=8$\,eV).

\begin{figure}[t]%
\centering%
\includegraphics[width=1.0\columnwidth]{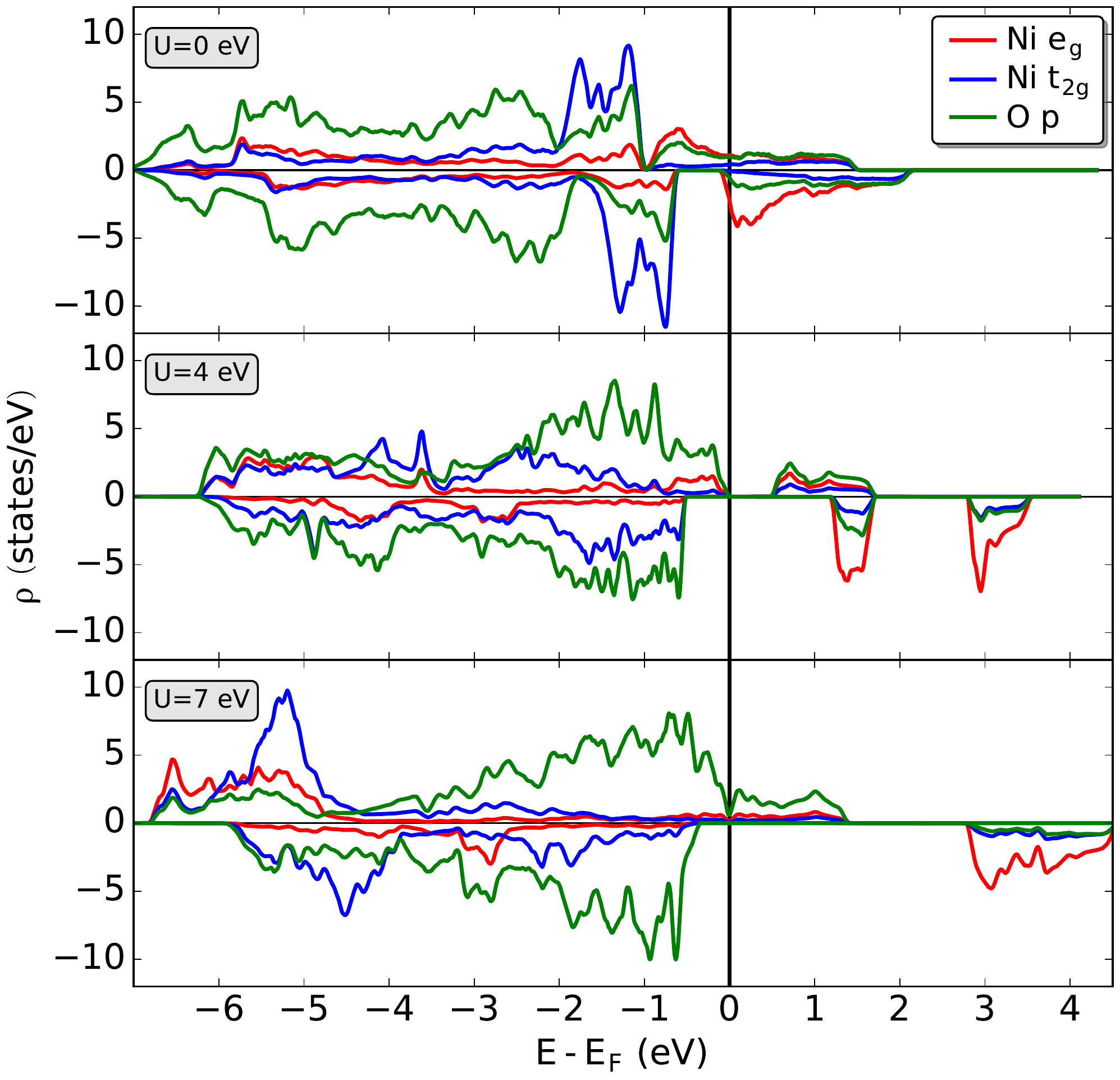}
\caption{Projected densities of states of fully relaxed LuNiO$_3$ with
  FM order for different $U$ values (and fixed $J=0$~eV). Only the Ni
  $d$ (separated in $t_{2g}$ and $e_g$ orbital character) and O $p$
  projections are shown (summed over all Ni and O sites within the
  unit cell). Minority spin states are depicted with negative sign and
  the Fermi level defines zero energy. One can see that with
  increasing $U$ the occupied Ni $d$-states, i.e., $t_{2g}$ (blue) and
  majority spin $e_g$ (red), are shifted down in energy relative to
  the oxygen $p$ states (green). For $U=7$~eV the occupied Ni $d$
  states are lower in energy than the oxygen $p$ states and the system
  becomes metallic.}
\label{fig:dos-collapse}
\end{figure}

The collapse of the breathing mode at large $U$ has also been observed
in earlier DFT calculations for NdNiO$_3$ by
\citeauthor{Prosandeev:2012fo}.~\cite{Prosandeev:2012fo} It can be
related to a qualitative change in the electronic structure beyond a
certain $U$ value. This is illustrated in Fig.~\ref{fig:dos-collapse},
which shows projected densities of states (DOS) for relaxed LuNiO$_3$
with FM order for $U=0$, $U=4$\,eV, and $U=7$\,eV (in all cases with
$J=0$). Here, the element-resolved DOS are summed over all atoms of a
given type, i.e., the Ni DOS contains the contributions from both LB
and SB sites.

For $U=0$, the Ni $d$ states (red and blue) are situated just
above the oxygen $p$ states (green). The system is slightly metallic
and no breathing mode appears in the relaxed structure. With
increasing $U$, the occupied Ni $d$ states are pushed down in energy
relative to the oxygen $p$ states, and a gap opens between the top of
the valence band with predominant O $p$ character and the conduction
bands with strong Ni $d$ character. This is indicative of a charge
transfer insulator with strong hybridization between the ligand $p$
and transition metal $d$ states. This is also the regime that supports
the breathing mode in the relaxed structure. The site splitting between
the two nickel sites can be observed as two distinct peaks (at
energies of approximately 1.5\,eV and 3\,eV) in the unoccupied
minority spin Ni $e_g$ DOS for $U=4$\,eV (middle panel of
Fig.~\ref{fig:lu-vs-pr-rel}).

However, for $U=7$ eV, the occupied Ni $d$ states are pushed
completely below the oxygen $p$ states, i.e, the system has entered a
negative charge transfer regime. This leads to reduced hybridization
between O $p$ and Ni $d$ states, and the unoccupied part of the
majority spin states has now essentially pure O $p$ character, i.e.,
it now clearly corresponds to two ligand holes. Interestingly, this
regime does not support the breathing mode distortion, as seen from the
top panel of Fig.~\ref{fig:lu-vs-pr-rel}. Thus, it appears that the
bond disproportionation in the nickelates depends strongly on the
degree of hybridization between the Ni $d$ and O $p$ states and
requires a mixed character of the nominal Ni $e_g$ bands. On the other
hand, if the ``ligand hole'' character of the unoccupied states
becomes too dominant, the bond disproportionation becomes
unfavorable. This is very much in line with the interpretation of
``charge order'' in terms of hybridized Ni-centered $e_g$-like Wannier
functions, as discussed by
\citeauthor{Varignon:2017is}.~\cite{Varignon:2017is}


Finally, in the bottom panel of Fig.~\ref{fig:lu-vs-pr-rel}, we
compare the relative stability of the FM and $T$-AFM states for the
two ``end-members'' LuNiO$_3$ and PrNiO$_3$ in the fully relaxed
structures as a function of $U$ (and using $J=0$). Here, a negative
(positive) value indicates that the $T$-AFM (FM) state is
energetically favored. One can see that, while for LuNiO$_3$ the FM
state is more favorable than $T$-AFM over essentially the whole range
of $U$ (with a nearly vanishing energy difference for $U=0$), for the
case of PrNiO$_3$ the energy difference $E_{T\text{-AFM}} -
E_{\text{FM}}$ exhibits a nonmonotonous behavior with a minimum at
around $U=1$\,eV.  Most strikingly, the $T$-AFM state is favored in
PrNiO$_3$ for $U$ values up to $U \approx 3$\,eV. Thus, in the small
$U$ regime (below 3-4~eV) the $T$-AFM state becomes more favorable for
increasing size of the \textit{R} cation, i.e., when going from Lu to
Pr. This is consistent with the experimentally observed trend for the
magnetic ordering temperature. We point out that, even though here we
show only data for the two end members of the series, we have verified
that the corresponding trends evolve continuously throughout the
series.

The results for the energy difference between $T$-AFM and FM for
LuNiO$_3$ are similar to the ones presented in
Fig.~\ref{fig:lunio3-exp-u-comp} for the experimental structure,
although in the experimental structure the $T$-AFM state is lower in
energy than FM for $U < 2$\,eV. This is due to the small structural
differences between the experimental and relaxed structures, which
slightly shift the energetics of the different magnetic orderings.
Additionally, we note that in our calculations the $T$-AFM ordering is
found to be stable in LaNiO$_3$ within $Pbnm$ symmetry. This is in
agreement with a very recent theoretical work by
\citeauthor{2017arXiv170808899S},~\cite{2017arXiv170808899S} where it
is also shown that the stability of the $T$-AFM ordering disappears if
the correct $R\bar{3}c$ symmetry is considered. This shows that also
LaNiO$_3$ is very close to a transition between the breathing mode
phase with AFM ordering and the metallic $R\bar{3}c$ phase. Together
with the differences found for LuNiO$_3$ in the experimental and
relaxed structures, it also demonstrates that the energy differences
between different magnetic states are rather sensitive to small
changes in the underlying crystal structure, indicating a subtle
interplay between magnetism and structure in the rare-earth
nickelates.

\section{Summary}
\label{sec:summary}

We have presented a systematic DFT+$U$ study for the whole series of
perovskite structure rare-earth nickelates. Our goal was to assess if
and to what extent the structural and magnetic properties of these
compounds can be described within the DFT+$U$ approach. In order to
distinguish different structural distortions, we have used a
symmetry-based mode decomposition.  Based on this decomposition, the
transition from the metallic $Pbnm$ structure at high temperatures to
the insulating $P2_1/n$ structure at lower temperatures can mainly be
related to a single octahedral ``breathing mode'' corresponding to irrep $R_1^+$
of the cubic reference structure.

We find that essentially all structural parameters apart from this
$R_1^+$ mode amplitude are rather well described already within
nonmagnetic DFT calculations with \mbox{$U=0$}. In particular, this is
the case for the important $R_4^+$ mode describing the degree of
out-of-phase octahedral rotations around the orthorhombic $a$ axis,
which decreases strongly from $R$=Lu towards $R$=La. However, in order
to obtain a nonzero $R_1^+$ mode amplitude in agreement with the
experimentally observed $P2_1/n$ low temperature structures, both
magnetic order and a nonzero value of $U$ are required within the
calculations. Thereby, the obtained amplitudes of the breathing mode
strongly depend on the value of $U$ and also on the magnetic order
imposed in the calculation. For not too large $U$, the $R_1^+$
amplitude increases with increasing $U$ and it is significantly larger
for the more realistic $T$-AFM order than for the FM case. For the
case with $T$-AFM order, very good overall agreement with the
experimentally determined structures across the whole series is
achieved for $U=2$\,eV and $J=1$\,eV. Similar good agreement can also
be achieved for FM order using a larger $U$ value of around
5\,eV. However, if $U$ is further increased, and once the occupied Ni
$d$ states are pushed energetically below the O 2$p$ manifold, the
$R_1^+$ mode vanishes again and the system becomes metallic.

Both our calculations as well as the available experimental data
indicate a decrease of the $R_1^+$ amplitude across the series from
$R$=Lu towards $R$=Pr. This decrease seems to be somewhat weaker in
our computational results compared to experiment. Here, we note that,
in order to simplify the analysis, we have always compared results
obtained with the same values for $U$ and $J$ across the whole series.
However,the use of a constant $U$ value for the whole nickelate series
might not be fully appropriate. Considering the strong effect of $U$
on the $R_1^+$ amplitude, even a small decrease of $U$ from $R$=Lu
towards $R$=Pr would result in a noticeably stronger decrease of the
$R_1^+$ amplitude across the series. Since the octahedral rotations
($R_4^+$ and $M_3^+$ modes) decrease towards $R$=Pr, and thus the
hybridization between the Ni $d$ and O $p$ states increases,
potentially leading to enhanced screening, the correct $U$ value for
$R$=Pr could indeed be slightly smaller compared to $R$=Lu. Therefore,
in order to clarify how large (or small) these effects really are,
first principles calculations of $U$ across the series would be of
great interest.

On the other hand, it should also be noted that the available
experimental data is quite sparse. In particular, data for the
compounds in the middle of the series, i.e., for $R$=Gd and Sm, is
currently not available. Furthermore, an unexpectedly large breathing
mode amplitude has been reported for HoNiO$_3$ at $60$~K below the
MIT~\cite{Alonso:2001bs} and for DyNiO$_3$ at
$2$~K~\cite{Munoz:2009go} (see Fig.~\ref{fig:r1+-mode}), and
systematic measurements of the temperature dependence of the $R_1^+$
amplitude are also lacking. In particular, considering the strong
influence of the magnetic state on the $R_1^+$ amplitude obtained in
the calculation, it would be of interest whether there is a noticeable
change in the $R_1^+$ amplitude (or some other structural parameters)
when the nickelate compounds (with $R$ from Lu to Sm) undergo the
transition to the AFM phase. Indeed, some anomalies of the phonon
frequencies at the magnetic transition temperature have already been
observed in SmNiO$_3$ thin films using Raman
scattering.~\cite{PhysRevB.78.104101}

While the overall trends and orders of magnitude seem to be well
captured within the DFT+$U$ calculations, some deficiencies also
become apparent.  For example, the imposed $T$-AFM ordering, which is
compatible with the experimental data, is only energetically favored
(compared to the FM state) for a relatively small range of $U$
values. For the case of LuNiO$_3$ it is even hardly favored at all
(only for $U=0$ in the fully relaxed structure). Nevertheless, in the
small $U$ regime, the $T$-AFM state becomes more and more
energetically favored with increasing radius of the rare-earth cation
(see bottom panel of Fig.~\ref{fig:lu-vs-pr-rel}), consistent with the
experimentally observed trend of the magnetic ordering
temperature. Our calculations also show that the $T$-AFM order
generally couples much stronger to the breathing mode distortion than
the FM order.

Overall, we find that the best agreement with experimental
observations across the whole series, regarding both structure and
magnetic order, is achieved if a relatively small value of $U \approx
2$\,eV is used in the calculations. This is consistent with the work
of \citeauthor{Varignon:2017is}~\cite{Varignon:2017is} and in contrast
to what has been suggested by
\citeauthor{Prosandeev:2012fo}~\cite{Prosandeev:2012fo} However, one
should note that even for $U=2$\,eV, the stability of the FM state
seems to be overestimated, in particular for the small rare-earth
cations such as Lu.

To conclude, our results give a clear picture of the predictive
capabilities of the DFT+$U$ approach in the rare-earth nickelate
series, and also provide a solid starting point for the use of more
advanced computational methods, such as, e.g., DFT+DMFT. Furthermore,
they can also be used as reference for future experimental
investigations regarding the temperature dependence of the structural
parameters and trends across the series.

\begin{acknowledgments}
We are indebted to Marisa Medarde, Oleg Peil, Antoine Georges, Michael
Fechner, and Gabriele Sclauzero for helpful discussions. This work was
supported by ETH Zurich and the Swiss National Science Foundation
through Grant No. 200021-143265 and through NCCR-MARVEL. Calculations
have been performed on the PASC cluster ``M\"o{}nch'', the MARVEL
cluster ``Daint'', both hosted by the Swiss National Supercomputing
Centre, and the ``Euler'' cluster of ETH Zurich.
\end{acknowledgments}

\bibliography{nickelate-paper}

\begin{thebibliography}{49}
\bibitem [{foo()}]{footnote1}%
  \BibitemOpen
  \href@noop {} {}\bibinfo {note} {We note that, strictly speaking, the AFM
  order with the given wave-vector is not compatible with $P2_1/n$ space group
  symmetry, and thus the symmetry of the system is further lowered within the
  AFM phase. However, the resulting effect on the structure is very small and
  is not relevant in the present context.}\BibitemShut {Stop}%
\end{thebibliography}

\end{document}